\documentclass{emulateapj}       
       
\shorttitle{Signature of wind RS on GRB afterglow} \shortauthors{Pe'er \& Wijers}

\newcommand{\Hz}{\rm{\, Hz }}

\newcommand{\beq}{\begin{equation}}       
\newcommand{\eeq}{\end{equation}}       
\newcommand{\ba}{\begin{array}}       
\newcommand{\ea}{\end{array}}       
    
\newcommand{\E}{E_{53}}    
      
\newcommand{\M}{\dot{M}_{-6}} 
 
\renewcommand{\v}{{\emph v_{w,8}}} 
\renewcommand{\t}{{t}_{\star,6}}

\newcommand{\n}{n_{0,3}} 
\newcommand{\ns}{n_{R_0,-1.5}}
\newcommand{\rs}{R_{0,18}}
      
\newcommand{\ee}{\epsilon_{e,-1}}      
\newcommand{\eB}{\epsilon_{B,-2}}

\def \etal{{\it et al.~}}      
      
\begin{document}       
       
\title{The signature of a wind reverse shock in Gamma-ray bursts afterglows}       
     
\author{Asaf Pe'er\altaffilmark{1}\altaffilmark{2} and Ralph A.M.J. Wijers\altaffilmark{1}}  
\altaffiltext{1}{Astronomical institute ``Anton Pannekoek'', Kruislaan 403, 1098SJ Amsterdam, the Netherlands} 
\altaffiltext{2}{apeer@science.uva.nl} 
\begin{abstract}     
Explosions of massive stars are believed to be the source of a
significant fraction of gamma-ray bursts (GRBs). 
If this is indeed the case, then the explosion blast wave propagates
into a complex density structure, composed of a stellar wind bounded
by two shock waves - a wind reverse shock and a forward shock. 
As the explosion blast wave reaches $R_0$, the radius of the wind
reverse shock, it splits into two shock waves - a reverse and a
forward shock wave.
We show that the reverse shock thus produced is not strong, therefore
full analytical treatment is required in calculating its properties. 
We calculate the dynamics of the flow and the
evolution of the blast waves in all of the different stages. 
We show that the fluid Lorentz factor at $r>R_0$ is equal
to $0.725$ times the blast wave Lorentz factor as it reaches $R_0$,
and is time (and $r$) independent as long as the blast wave reverse
shock exists. 
Following the calculation of the blast wave evolution, we
calculate the radiation expected in different energy bands.
We show that about a day after the main explosion,
as the blast wave reaches $R_0$, the observed afterglow flux starts to 
rise. It rises by a factor of about 2 in a few hours, during
which the blast wave reverse shock exists, and then declines.
We show that the power law index describing light curve time evolution
is different at early (before the rise) and late times, and is
frequency dependent.  
We present light curves in the different energy bands for this
scenario.     
\end{abstract}

\section{Introduction}
\label{sec:intro}
In recent years, there is increasing evidence that long duration
\citep[$t_{90}\geq$ 2 s;][]{K93} gamma-ray bursts (GRBs) are associated
with the deaths of massive stars, presumably arising from core collapse
\citep{Woosley93, LE93, MW99, MWH01, ZWM03}. This evidence includes the
association of some GRBs with type Ib/c supernovae [ GRB980425 and
SN1998bw \citep{G98}; GRB011121 and SN2001ke \citep{Gar03}; GRB030329
and SN2003dh \citep{Hjorth03, Stanek03}; and GRB031203 and SN2003lw
\citep{Malesani04}], as well as the association of GRBs with massive
star forming regions in distant galaxies \citep{Pac98, Wijers98,
  Fruchter99, TRRB02}. Further clues arise from evidence of high
column densities toward GRBs, which associate GRBs with molecular
clouds \citep[e.g.,][]{GW01}.

If indeed GRBs are associated with the deaths of massive stars, then
the circumburst environment is influenced by the wind from
the star. As supersonic wind from the star meets the interstellar
medium (ISM), two shock waves are formed: a forward shock wave
that propagates into the ISM, and a reverse shock that propagates into
the wind (in the wind rest frame). The ``wind bubble'', thus formed is
composed of the unshocked wind, the shocked wind and the shocked ISM
\citep{CMW75, WMCSM77}. This complex structure of the circumburst
environments is expected to affect the dynamics of the GRB blast wave
during its late (afterglow) evolution, hence to have an observable effect
on the afterglow emission \citep{Wijers01, RRDMT01, CLF04, RRGSP05,
  EGDM05}. 

The main effect is expected to take place when the relativistic blast
wave reaches the density discontinuity produced by the wind reverse shock.
The blast wave then splits into two shock waves - a blast wave forward
shock and a blast wave reverse shock \citep{SP95, RRGSP05}.  
The blast wave reverse shock thus produced, propagates into a
wind that was already shocked by the original blast wave prior to its
split. This wind is hot, therefore the ratio of the energy
densities, or of the gas pressures downstream and upstream of the flow
past the blast wave reverse shock is not much greater than unity,
hence the blast wave reverse shock is not strong.
Obviously, as the blast wave splits into two, the dynamics of the
created shock waves can no longer be described by a self-similar
motion \citep{BM76}, which determines the
evolution of the original relativistic blast wave at earlier stages,
as well as the dynamics at a much later stage. 

In this paper, we analyze in detail the effect of the circumburst
environment on the dynamics of the blast wave(s). Following the
analysis of \citet{CMW75}, we determine in \S\ref{sec:environment}
expected circumburst conditions for a GRB progenitor.
In \S\ref{sec:shock_collision} we use the jump conditions at the blast
wave reverse and forward shocks to show that a simple analytic relation
between the velocity of the shocked fluid prior to the blast wave split
and the velocity of the shocked fluid after the split is obtained.
We further find a simple analytical relation that connects the
Lorentz factor of the reverse shock to the Lorentz factor of the fluid
as it reaches the contact discontinuity. We use
these results to determine in \S\ref{sec:shock_dynamics} the evolution
of the fluid velocity. We calculate in \S\ref{sec:lightcurve} the
resulting light curves in different energy bands, before summarizing
and discussing the implications of our findings in \S\ref{sec:summary}.

\section{The circumburst environment}
\label{sec:environment}

The circumburst environment during the GRB explosion depends on the
evolutionary stages of the progenitor, prior to its final (presumably,
Wolf-Rayet) phase. A standard evolutionary track for a massive galactic star
is to start as an O star, evolve through a red supergiant (RSG) phase
or luminous blue variable (LBV) phase, before ending as a Wolf-Rayet
star \citep{GML96,GLM96}. The RSG phase may be absent for low
metallicity stars \citep{Chieffi03}, which are preferred as GRB
progenitors \citep{LeFloch03, Fynbo03, Vreeswijk04}, or for rapidly
rotating stars \citep{Pet05}. 

We thus consider a massive ($M\gtrsim 25 M_\odot$), low metallicity
($Z\sim 0.01$) star as a GRB progenitor. 
During the Wolf-Rayet phase of the star, which lasts a duration of
$\sim 10^6$~yr, the star loses mass at 
a typical mass loss rate of $\dot M \approx 10^{-6} \, M_\odot {\rm
  yr^{-1}}$, producing a wind characterized by a typical velocity $v_w
\sim 1000 {\rm\, km\, s^{-1}}$, presumably steady during
most of the Wolf-Rayet phase of the star \citep{VKL00, CLF04, VK05}.  
The evolution of the wind-driven circumstellar shell was first derived
by \citet{CMW75} and \citet{WMCSM77}. It was shown that during most of 
its lifetime (neglecting very short early stages), the system has four
zones consisting, from the inside out, of ($a$) a hypersonic stellar
wind with characteristic density $n_a(r) = \dot M /(4 \pi m_p r^2
v_w)$; 
($b$) a hot, almost isobaric region consisting of shocked stellar wind
mixed with a small fraction of swept-up interstellar gas; ($c$) a
thin, dense shell containing most of the swept-up interstellar
gas; ($d$) ambient interstellar gas.

Neglecting the width of region ($c$) compared to region ($b$) (see below),
and assuming that most of the energy in region ($b$) is in the form of
thermal energy, it was shown by \citet{CMW75} that the outer
termination shock radius is at
\beq
\ba{lll}
R_{fs,wind} & = & \left({125 \over 308 \pi}\right)^{1/5} \left( {\dot M
  v_w^2 t_\star^3 \over \rho_{ISM}}\right)^{1/5} \nonumber \\
& = & 1.6 \times 10^{19} \, \M^{1/5} \v^{2/5}
\n^{-1/5} \t^{3/5} {\rm \, cm}, 
\ea
\eeq   
 where $\dot M = 10^{-6} \M \, M_\odot {\rm yr^{-1}}$, $v_w = 10^8 \v
 {\rm \, cm \, s^{-1}}$, the ambient ISM density is $\rho_{ISM}=m_p
 n_{ISM}$,  $n_{ISM} = 10^3 \, \n {\rm \, cm^{-3}}$ and  $t_\star =
 10^6 \t {\rm \, yr}$ is the lifetime of the Wolf-Rayet phase of the star. 
Here we have scaled the ambient density to a value typical of a
molecular cloud, in which we assume the young star to be still embedded.  
The density of the swept up ISM matter in region ($c$) is approximated
by its value for a strong, adiabatic shock, $\rho_c \simeq 4 \rho_{ISM}$. 
\footnote{Detailed models of shock propagation into ambient medium
\citep[e.g.,][]{CLF04,EGDM05} suggest that the numerical pre-factor
depends on early stages of the stellar evolution, and may be different
than 4. Nonetheless, we show in \S\ref{sec:shock_collision} that the
explosion blast wave is not expected to reach this region while
relativistic, and therefore the exact value of the density in this region
has no observational consequences.}    
Comparison of the total ISM mass swept up to radius
$R_{fs,wind}$, $M_{ISM} \simeq (4\pi/3)R_{fs,wind}^3 \rho_{ISM}$ to
the mass in region ($c$), $\approx 4 \pi R_{fs,wind}^2 \Delta R_c
\rho_c$ then leads to the conclusion that the width of region ($c$) is
$\Delta R_c \approx R_{fs,wind}/12$.  

The pressure in regions ($b$) and ($c$) is $P_b = P_c = (2/3)
 u_b$, (assuming a monatomic gas) where $u_b$ is the energy 
 density in region ($b$), or
\beq
\ba{lll}
P_b = P_c & = & {7 \over 25}\left({125 \over 308 \pi}\right)^{2/5}  \rho_{ISM}
\left({\dot M v_w^2 \over \rho_{ISM} t_\star^2}\right)^{2/5} \nonumber \\
& = & 1.4 \times 10^{-10} \,\M^{2/5} \v^{4/5} \n^{3/5} \t^{-4/5} {\rm
 \, dyn \, cm^{-2}}.   
\ea
\eeq
The radius of the inner (reverse) shock was calculated by
\citet{WMCSM77} \citep[see also][]{GF96}, assuming that the pressure in
region ($b$) is much larger than the pressure in region ($a$) $P_b \gg
P_a$ (strong shock assumption), by equating the momentum flux upstream
and downstream of the shock. Comparison of the ram pressure in the upstream
region ($a$), $\rho_a(R_0) v_w^2$ to the pressure downstream, $P_b +
\rho_b v_b^2$, leads to
\beq
\ba{lll}
R_0 \equiv R_{rs,wind} & = & \left({3 \over 4}{\dot M v_w \over 4 \pi
  P_b}\right)^{1/2} \nonumber \\
& = & 1.6 \times 10^{18} \, \M^{3/10} \v^{1/10}
\n^{-3/10} \t^{2/5} {\rm \, cm}
\label{eq:R_0}
\ea
\eeq  
where $\rho_a(R_0) = \dot M /(4 \pi R_0^2 v_w)$, and $\rho_b = 4
\rho_a$, $v_b = v_w/4$ (strong shock assumptions) were used.
The number density of particles in region ($a$) is $n_a(r) \propto
r^{-2}$, and at $r=R_0$ is given by
\beq
 n_a(R_0) = {\dot M \over 4 \pi m_p R_0^2 v_w} = 3.0 \times 10^{-2} \,
\rs^{-2} \M \v^{-1} {\rm \, cm^{-3}},
\label{eq:n_a} 
\eeq
where $R_0 = 10^{18} R_{0,18} {\rm \, cm}$.\footnote{Note that this
  equation differs from \citep{Wijers01} due to an error in the latter.}
The density in region ($b$) depends on the uncertain physics of the
heat conduction. Heat conduction could be prevented by a magnetic
field, which is expected to be toroidal in this region
\citep{CLF04}. Under this assumption, and using the fact that 
the internal speed of sound in this region is much higher than the
expansion velocity, the number density in region ($b$) is
approximately $r$~-independent, and is equal to $n_b \simeq 4 n_a(R_0)$
\citep{WMCSM77}; \citep[see also][]{DW97}.

\begin{figure}          
\plotone{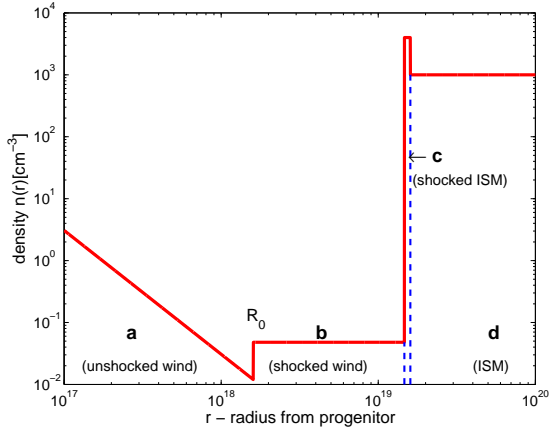}          
\caption{Schematic density profile for the scenario of a massive star
  emitting wind. Region ($a$) consists of the unshocked wind, with a
  density profile $n(r) \propto r^{-2}$. Region ($b$) consists of hot,
  nearly isobaric shocked wind. Region ($c$) consists of the shocked
  ISM, and region ($d$) is the unshocked ISM.    
}          
\label{fig:density}          
\end{figure}

A schematic density profile of the bubble is shown in Figure
\ref{fig:density}. While being only a schematic representation, the
density profile presented is in very good agreement with detailed
models of stellar evolution \citep{CLF04,RRGSP05,EGDM05}.
       
\section{Blast wave interaction with the density discontinuity at $R_0$}
\label{sec:shock_collision}
  
We now consider the relativistic blast wave created by the explosion 
producing the GRB. Since the rest-mass energy of the material in
region ($b$), $E_{RM} = \dot M t_\star c^2 \approx 2\times 10^{54} \M
\t {\rm\, erg}$ is larger than the isotropically equivalent energy
released in the explosion producing the GRB, $\sim 10^{53} {\rm\,
  erg}$, the blast wave cannot cross region ($b$) while
relativistic. Therefore, all the observable effects are 
expected to occur as the blast wave propagates in regions ($a$) and ($b$). 
In region ($a$), at radius $r < R_0$, the blast wave evolution is well 
approximated by the \citet{BM76} self-similar evolution for an
explosion into a density gradient. A similar description holds in region
($b$) for $r\gg R_0$ (as long as the blast wave remains
relativistic). We thus concentrate on the interaction of the blast
wave with the density discontinuity at $R_0$.

\begin{figure}          
\plottwo{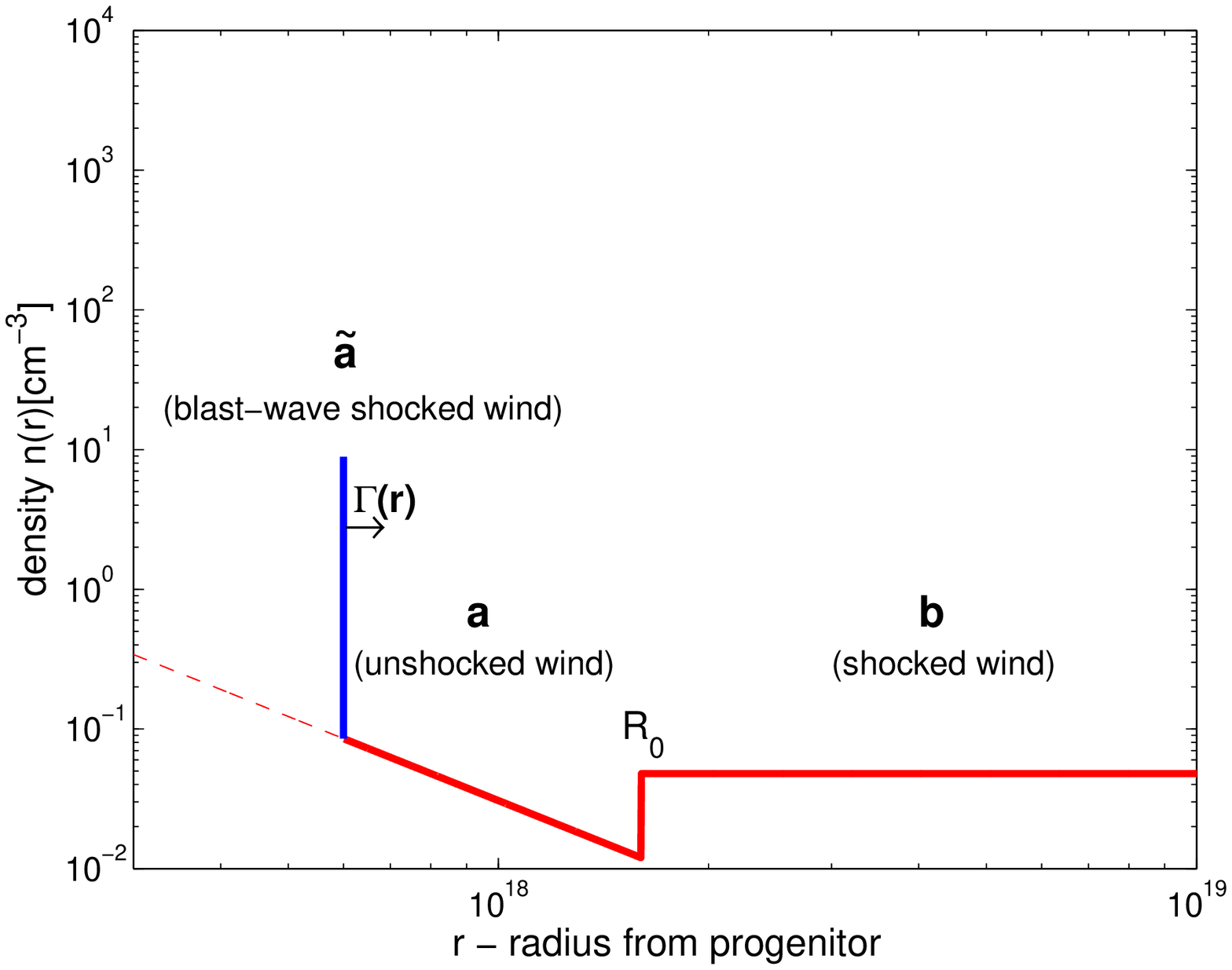}{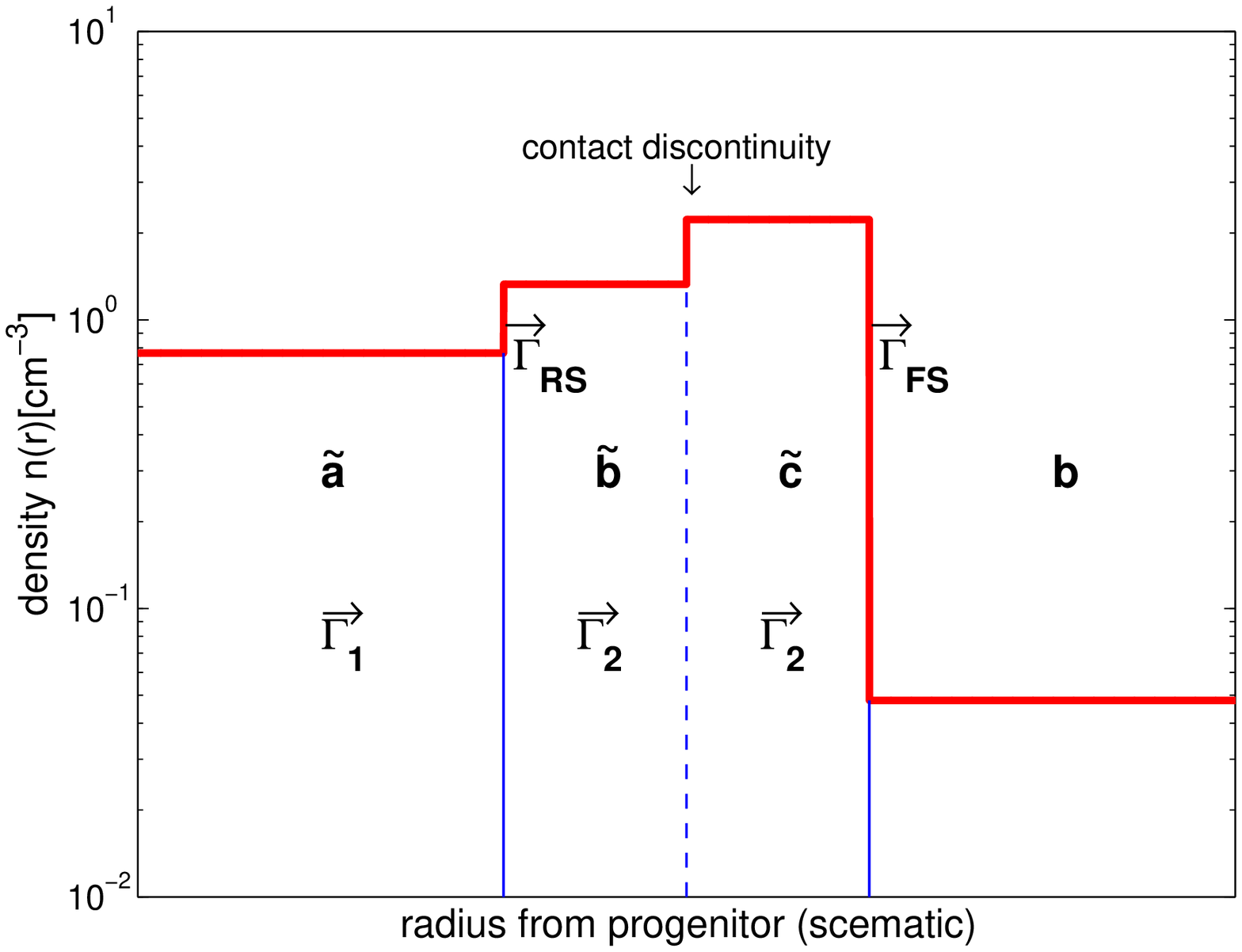}      
\caption{Schematic description of the different regimes during the
  blast wave evolution. Left: blast wave evolution in region ($a$)
  results in the creation of region $(\tilde a)$, whose observed width
  $\Delta R \approx r/(4 \Gamma^2) \sim 10^{15} {\rm \, cm}$ is too
  narrow to be observed on the scale of the plot. Right: schematic
  description of the different plasma regimes as the blast wave
  propagates at $R_0<r<R_1$, and the reverse shock exists. Region
  $(\tilde b)$ contains plasma from region $(\tilde a)$ shocked by the
  reverse shock, and region  $(\tilde c)$ contains plasma from region
  $(b)$ shocked by the forward shock. $\Gamma_1 > \Gamma_2 >
\Gamma_{RS}$. \\
\newline
}          
\label{fig:density_profile_with_shock}          
\end{figure}

Consider a relativistic blast wave that propagates in region
($a$). The matter in the downstream region of the shock, which we
denote as region $(\tilde a)$, is composed of the shocked material of
region ($a$). Being shocked by a relativistic shock wave its
temperature is much higher than $m_ec^2$, thus the pressure in this
region is related to the energy density in the region by the relation
$P_{\tilde a} = (1/3) u_{\tilde a}$ (as opposed to the relation used
in \S\ref{sec:environment} describing the flow in region ($b$), $P_b=
(2/3) u_b$, which is valid  
for flow with temperature much smaller than $m_ec^2$, as is the case
in region ($b$) prior to the blast wave propagation). 
Since for $r<R_0$ the fluid in region $(\tilde a)$ is downstream from
the flow past the shock wave, it thermalizes, hence isotropizes
immediately after passing the shock. 
We therefore adopt the commonly used approximation that the fluid in
region  $(\tilde a)$ is uniform and isotropic.
Under this assumption, its number and energy densities are 
$n_{\tilde a}(r) \simeq 4 \Gamma(r) n_a(r)$, 
$u_{\tilde a}(r) \simeq 4 \Gamma^2(r) n_a(r) m_p c^2$ respectively, 
where $\Gamma(r)$ is the Lorentz factor of the flow downstream (see
figure \ref{fig:density_profile_with_shock}, left), and $\Gamma(r) \gg
1$ assumed  \citep{BM76}.

As the relativistic blast wave reaches $R_0$, region ($a$) no longer
exists, and all of its matter is swept up by the shock and is in
region $(\tilde a)$. At $R_0$ the blast wave splits into two: a
relativistic forward shock that continues to propagate forward into
the matter in region ($b$), and a reverse shock that propagates into
region $(\tilde a)$. Thus, two new regions are formed: region $(\tilde
b)$ which contains matter from region $(\tilde a)$ shocked by the reverse
shock, and region $(\tilde c)$ which contains matter from region ($b$)
shocked by the forward shock. Regions $(\tilde b)$ and  $(\tilde c)$
are separated by a contact discontinuity. The fluids in regions
$(\tilde b)$ and $(\tilde c)$ are both at rest relative to the contact
discontinuity. Therefore, the fluids in these two regions propagate at the
same Lorentz factor $\Gamma_2$ (in the observer frame),
which is smaller than $\Gamma_1$ - the Lorentz factor of the fluid in
region ($\tilde a$) at $r=R_0$ (figure
\ref{fig:density_profile_with_shock}, right).  
 
While the relativistic forward shock is strong, the reverse shock thus
produced cannot be strong: for a relativistic forward shock, the
energy density in region $(\tilde c)$ is $u_{\tilde c} \simeq 4
\Gamma_2^2 \omega_b$, where $\omega_b \simeq n_b m_p c^2$ is
the enthalpy in region ($b$)\footnote{The temperature in region ($b$)
is $T_b \simeq  10^7 \v^2$~K (neglecting radiative cooling) and may be
lower if radiative cooling is considered \citep{CMW75,RRGSP05}. 
Therefore the thermal energy in this region is much smaller then the
rest-mass energy, and the pressure is much smaller than the
relativistic energy density.}. 
For a strong reverse shock, the energy density in region $(\tilde b)$
is given by $u_{\tilde b} \simeq (4 \bar{\Gamma_2}+3)\bar{\Gamma_2}
\omega_{\tilde a}$. Here, $\omega_{\tilde a} = (4/3) 4 \Gamma_1^2
n_a(R_0) m_p c^2$ is the enthalpy in region $(\tilde a)$, and
$\bar{\Gamma_2}$ is the Lorentz factor of the fluid in region $(\tilde 
b)$ as viewed in the rest frame of region $(\tilde a)$,
$\bar{\Gamma_2} \simeq (1/2)(\Gamma_1/\Gamma_2 +\Gamma_2/\Gamma_1)$.
The term $+3$ in the expression for $u_{\tilde b}$ is added because the
Lorentz factor of the fluid in region $(\tilde 
b)$ as viewed in the rest frame of region $(\tilde a)$
$\bar{\Gamma_2}$, is only mildly relativistic.
Equating the energy densities at both sides of the contact
discontinuity separating regions $(\tilde b)$ and $(\tilde c)$,
using $n_b = 4 n_a(R_0)$, leads to $\Gamma_2^2 =
\bar{\Gamma_2}[(4/3)\bar{\Gamma_2}+1]\Gamma_1^2$. Since 
$\bar{\Gamma_2}>1$, the requirement $\Gamma_2 < \Gamma_1$ cannot be
fulfilled. We therefore conclude that the reverse shock formed as the
blast wave reaches $R_0$ is not strong.
 
The Lorentz factor $\Gamma_2$ of the fluid in regions $(\tilde b)$ and
$(\tilde c)$, the number density in region~$(\tilde b)$, $n_{\tilde
  b}$, and the Lorentz factor of the blast-wave reverse shock,
$\Gamma_{RS}$, are found using the reverse shock jump conditions and
the requirement of pressure balance across the contact discontinuity,
which leads to $u_{\tilde b}= u_{\tilde c}$. 
We write the Taub adiabatic \citep[e.g.,][]{LL59} at the reverse shock
in the form 
\beq
{n_{\tilde b}^2 \over n_{\tilde a}^2} = {\omega_{\tilde b} \over
  \omega_{\tilde a}} \left( {\omega_{\tilde b} - P_{\tilde b} +
  P_{\tilde a}} \over {\omega_{\tilde a} + P_{\tilde b} -
  P_{\tilde a}} \right) = {4 \Gamma_2^2 \over \Gamma_1^2} \left(
  {\Gamma_1^2 + 12 \Gamma_2^2 \over 3 \Gamma_1^2 + 4\Gamma_2^2}
  \right) .
\label{eq:Taub}
\eeq
Here, $P_{\tilde a} = u_{\tilde a}/3$,  $P_{\tilde b} = u_{\tilde
  b}/3$ are the pressures in regions $(\tilde a)$,$(\tilde b)$ (the
temperature in region $(\tilde b)$ is higher than the temperature in
region $(\tilde a)$, which is much higher than $m_e c^2$, thus
relativistic formulae are used), 
$ u_{\tilde b} = u_{\tilde c} = 4 \Gamma_2^2 n_b m_p c^2$ is the
energy density in region $(\tilde b)$, $u_{\tilde a} = 4 \Gamma_1^2
n_a(R_0) m_p c^2$ is the energy density in region $(\tilde a)$, the
enthalpies in regions $(\tilde a)$, $(\tilde b)$ are $\omega_{\tilde a}=
u_{\tilde a} + P_{\tilde a}$, $\omega_{\tilde b}= u_{\tilde b} +
P_{\tilde b}$ respectively, and we have used $n_b = 4 n_a(R_0)$ in the
derivation of the second equality.  

By making a Lorentz transformation from the reverse shock rest frame
to the observer frame, conservation of particle number flux at
the reverse shock is written in the form
\beq
\beta_{RS} = {n_{\tilde a} \Gamma_1 \beta_1 - n_{\tilde b} \Gamma_2
  \beta_2 \over n_{\tilde a} \Gamma_1 -  n_{\tilde b} \Gamma_2} = {
  {(n_{\tilde a} / n_{\tilde b})} \Gamma_1 \beta_1 - \Gamma_2
  \beta_2 \over {(n_{\tilde a} / n_{\tilde b})} \Gamma_1 - \Gamma_2},
\label{eq:beta_rs}
\eeq 
where $\beta_{1,2} \equiv (1-1/\Gamma_{1,2}^2)^{1/2}$ is the
normalized velocity of the fluid in regions $(\tilde a)$,$(\tilde b)$
respectively, and $\beta_{RS} \equiv (1-1/\Gamma_{RS}^2)^{1/2}$ is the
normalized velocity of the reverse shock in the observer frame. 
As the third equation we use the continuity of the energy flux at the
reverse shock, which after Lorentz transformation to the observer
frame, takes the form
\beq
\omega_{\tilde a} \Gamma_1^2 \left( 1-\beta_1
\beta_{RS}\right)(\beta_1-\beta_{RS}) 
= \omega_{\tilde b} \Gamma_2^2 \left( 1-\beta_2
\beta_{RS}\right)(\beta_2-\beta_{RS}). 
\label{eq:energy_flux}
\eeq 

Equations \ref{eq:Taub}, \ref{eq:beta_rs} and \ref{eq:energy_flux},
which connect the thermodynamic properties upstream and downstream of
the flow past the reverse shock are sufficient to calculate the
unknown values of the thermodynamic variables $\Gamma_2$,
$\Gamma_{RS}$ and $n_{\tilde b}$.  
The calculation is done in the following way: 
inserting $\beta_{RS}$ from equation \ref{eq:beta_rs} and using
$\omega_{\tilde b}/\omega_{\tilde a} =4\Gamma_2^2/\Gamma_1^2$, 
equation \ref{eq:energy_flux} takes the form
\beq
\ba{l}
\Gamma_1^2 \left[ \left({n_{\tilde a} \over n_{\tilde b}}\right) +
  \Gamma_1 \Gamma_2 (\beta_1 \beta_2 -1 ) \right] \nonumber \\ 
= 4 \Gamma_2^2 \left[ \left({n_{\tilde a} \over n_{\tilde b}}\right) 
 \Gamma_1 \Gamma_2 (1 - \beta_1 \beta_2 ) - 1 \right] 
\left({n_{\tilde a} \over n_{\tilde b}}\right). 
\label{eq:energy_flux2}
\ea
\eeq 
Equation \ref{eq:energy_flux2} is simplified by approximating
$\beta_{1,2} \approx 1- 1/(2\Gamma_{1,2}^2)$, which leads to 
\beq
2 \Gamma_1 \Gamma_2 \left({n_{\tilde a} \over n_{\tilde b}}\right)
\left( \Gamma_1^2 + 4 \Gamma_2^2 \right) = \left( \Gamma_1^2 +
\Gamma_2^2 \right) \left[ \Gamma_1^2 + 4 \Gamma_2^2 
\left({n_{\tilde a} \over n_{\tilde b}}\right)^2 \right].
\label{eq:energy_flux3}
\eeq
Inserting the ratio $n_{\tilde a}/n_{\tilde b}$ from equation
\ref{eq:Taub} into equation \ref{eq:energy_flux3}, after some algebra we
are left with a quadratic 
equation for $\Gamma_2^2$, 
\beq
32 \Gamma_2^4 + 8 \Gamma_1^2 \Gamma_2^2 -13 \Gamma_1^4 = 0,
\eeq
with a solution
\beq
\ba{lcr}
\Gamma_2^2 & = & {\sqrt{27} -1 \over 8} \Gamma_1^2, \nonumber \\
\Gamma_2 & \simeq & 0.725 \Gamma_1.
\label{eq:Gamma2}
\ea  
\eeq
Inserting this result in equation \ref{eq:Taub}, the number density in
region $({\tilde b})$ is given by
\beq
{n_{\tilde b}^2 \over n_{\tilde a}^2} = {153 \sqrt{3} -259 \over 2}
\approx 3, 
\label{eq:n_tildeb}
\eeq
or $n_{\tilde b} \simeq 1.73 n_{\tilde a}$.
The energy density in region  $({\tilde b})$ is
\beq
{\omega_{\tilde b} \over \omega_{\tilde a}} = 
{u_{\tilde b} \over u_{\tilde a}} = { 4\Gamma_2^2 \over \Gamma_1^2} =
{3 \sqrt{3} -1 \over 2} \approx 2.1, 
\label{eq:u_tildeb}
\eeq
which means that the energy per particle in region $({\tilde b})$ is 
$u_{\tilde b}/n_{\tilde b} \simeq (2.1/1.73) u_{\tilde a}/n_{\tilde
  a} \approx 1.21 u_{\tilde a}/n_{\tilde a}$. We therefore conclude
that the energy per particle is increased by $\sim 20\%$ as the
particle passes through the reverse shock, from region  $({\tilde a})$
to  $({\tilde b})$.  
The Lorentz factor of the reverse shock is calculated using equation
\ref{eq:beta_rs}, 
\beq
\Gamma_{RS} \simeq 0.43 \Gamma_1.
\label{eq:Gamma_RS}
\eeq

\section{Blast wave evolution}
\label{sec:shock_dynamics}

As long as the explosion blast wave propagates in region $(a)$ at
$r<R_0$, its evolution is well described by the \citet{BM76}
self-similar solution for an explosion into a density gradient $n(r)
\propto r^{-2}$,
\beq
\Gamma(r; r<R_0) = \left( {9 E \over 16 \pi A c^2} \right)^{1/2} {1 \over
  r^{1/2}},
\label{eq:Gamma_r_small}
\eeq
where $E$ is the (isotropically equivalent) explosion energy, and
$A\equiv \dot M/(4 \pi v_w)$.
At $r=R_0$,
\beq
\Gamma_1 \equiv \Gamma(r=R_0) = 20.5 \, \E^{1/2} \ns^{-1/2} \rs^{-3/2},
\eeq
where $n_a(R_0) = 10^{-1.5} n_{R_0,-1.5} {\rm \, cm^{-3}}$.

The spatial dependence of the hydrodynamic variables of a shocked
fluid element (in region $(\tilde a)$), are given by the \citet{BM76}
self similar solution. In this solution, as the blast wave expands to 
radius $r$, more than 90\% of the energy and the particles are
concentrated in a shell of co-moving thickness   
$\Delta  r_{\tilde a}^{co.} = \xi r / \Gamma(r)$, where $\xi$ is
 a numerical factor, in the approximate range $0.1 - 0.5$ for the
hydrodynamical quantity in question (number density, energy, Lorentz
 factor etc.). 
Adopting the approximation that the fluid in region $(\tilde a)$ is
homogeneously distributed, we write  
the comoving width of this region at $r=R_0$ as 
$\Delta  R_{\tilde a}^{co.}(r=R_0) = R_0 / 8 \zeta \Gamma_1$,
where $\zeta$ is a numerical factor of order unity, which is
inserted in order to parametrize the discrepancy between the actual
density and energy profiles and the homogeneity approximation used.  
The Lorentz factor of each fluid element in region $(\tilde a)$ at
$r=R_0$ is therefore approximated to be $\Gamma(r=R_0)=\Gamma_1$.  

At $r>R_0$, the flow in region $(\tilde a)$ which was downstream the
flow past the blast wave at $r<R_0$, becomes upstream the flow past
the reverse shock. 
A fluid element in region $(\tilde a)$ therefore continues to move at
an approximately constant ($r$-independent) Lorentz factor
$\Gamma=\Gamma_1$, as long as the reverse shock exists
\footnote{The velocity of the reverse shock in the rest frame
  of region $(\tilde a)$, $\bar{\beta_{RS}} = 0.69$, is of course
  larger than the speed of sound, $\beta_{sound} = 0.57$. We therefore
do not expect a significant change in the thermodynamic properties of
  region  $(\tilde a)$ due to adiabatic expansion during the reverse
  shock crossing time.}. 
The shock jump conditions analyzed in \S\ref{sec:shock_collision}
then imply that the Lorentz factor of a fluid element in regions
$(\tilde b)$ and $(\tilde c)$ is also $r$-independent, given by
equation \ref{eq:Gamma2}, $\Gamma_2 = 0.725 \Gamma_1 = 14.8 \,
  \E^{1/2} \ns^{-1/2} \rs^{-3/2}$.  

The Lorentz factor of the reverse shock during its lifetime is
determined by the reverse shock jump conditions as well, hence is
$r$-independent and given by equation \ref{eq:Gamma_RS}, $\Gamma_{RS} =
0.43 \Gamma_1 = 8.8\, \E^{1/2} \ns^{-1/2} \rs^{-3/2}$. 
The reverse shock therefore completes its crossing through region
$(\tilde a)$ at distance 
\beq
R_1 = R_0 + { \Delta R_{\tilde a}^{ob.}(r=R_0) \over \beta_{1} -
  \beta_{RS}} \simeq R_0 \left( 1 + {1 \over 17.6 \zeta} \right) = 1.06 R_0
\eeq
from the explosion.
Here, $\Delta R_{\tilde a}^{ob.}(r=R_0) = \Delta R_{\tilde
  a}^{co.}(r=R_0)/\Gamma_1$ is the observer frame width of region
$(\tilde a)$ at $r=R_0$, and $\zeta=1$ assumed in the last
equality. At $r>R_1$ region $(\tilde a)$ no longer exists as
all of its content is in region  $(\tilde b)$, and the reverse shock
disappears. 

The total mass of the matter swept up by the blast wave as it
propagates from $R_0$ to $R_1$ is   
\beq
\ba{lll}
m_b(r = R_1) & = & {4 \pi \over 3} m_p n_b \left( R_1^3 -R_0^3 \right)
\nonumber \\
& =& {4 \over 3}\left[ \left( {R_1 \over R_0} \right)^3 -1 \right] m_a
\simeq 0.24 m_a, 
\label{eq:m_b}
\ea
\eeq 
where $m_a = 4 \pi m_p \int_{r=0}^{R_0} n_a(r) r^2 dr$ is the total
swept up mass of region $(a)$.   
The thermal energy of the particles swept up as the blast wave
propagates from $R_0$ to $R_1$, $E_{th} \simeq \Gamma_2^2 m_b c^2$
(assuming no radiative losses) is much larger than the fluids' kinetic
energy at $R_1$, $E_k \simeq \Gamma_2 (m_a + m_b) c^2$. 
As long as the reverse shock exists, the energy excess is
compensated by kinetic energy loss of particles moving through the
reverse shock, from region  $(\tilde a)$ to region  $(\tilde b)$. 
At $r>R_1$ the reverse shock no longer exists, and therefore the
plasma decelerates. Since at this stage the thermal energy is already
much larger than the kinetic energy, a self similar expansion follows
soon after the reverse shock ceases to exist, at $r \simeq R_1$. 
This self-similar expansion is well described by the \citet{BM76}
solution for an expansion into a uniform density medium. We thus
conclude that at $r>R_1$ the fluid Lorentz factor is given by
$\Gamma(r>R_1) = \Gamma_2 (r/R_1)^{-3/2}$. 

\begin{figure}          
\plotone{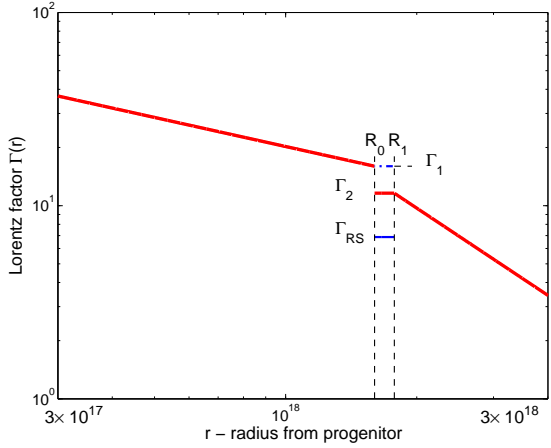}          
\caption{Schematic description of the plasma Lorentz factor as a
  function of $r$. For $r<R_0$, $\Gamma(r) \propto r^{-1/2}$; 
  At $R_0< r<R_1$, plasma in region $(\tilde a)$ continues to move at
  $\Gamma_1 \equiv \Gamma(r=R_0)$, while plasma in regions $(\tilde
  b)$, $(\tilde c)$ moves at Lorentz factor $\Gamma_2 = 0.725
  \Gamma_1$. The reverse shock moves at Lorentz factor $\Gamma_{RS} =
  0.43 \Gamma_1$. At $r>R_1$, plasma in regions $(\tilde b)$, $(\tilde
  c)$ moves at a self-similar motion with Lorentz factor $\Gamma
  \propto r^{-3/2} $.  
}          
\label{fig:Lorentz_factor}          
\end{figure}

The evolutionary stages of the fluid's Lorentz factor in the different
regimes are summarized in figure \ref{fig:Lorentz_factor}. The initial
blast wave exists at $r<R_0$. At $R_0<r<R_1$ fluid in region $(\tilde
a)$ moves at a constant Lorentz factor $\Gamma_1$, while fluid in
regions $(\tilde b)$ and  $(\tilde c)$ moves at a constant Lorentz
factor $\Gamma_2$. At $r>R_1$, the fluid occupies
regions $(\tilde b)$ and $(\tilde c)$ only, and moves at a self similar
motion. The blast wave becomes non-relativistic ($\Gamma_2-1 \simeq
1$) at radius $R_{NR} \simeq 3.3 R_1 \simeq 3.5 \times 10^{18} \, \rs
{\rm \, cm}$.

\section{Afterglow Re-brightening}
\label{sec:lightcurve}

The complex dynamic of the blast wave evolution has observational
consequences. In this section, we calculate the expected light curve in
this scenario. In our calculations, we use the standard synchrotron
emission model, which is in very good agreement with afterglow
observations \citep[e.g.,][]{WRM97,vPKW00}. 
We divide the calculation of the emitted radiation into the 3 different
phases, corresponding to the 3 phases of the blast wave evolution:
(a) the earliest phase, which corresponds to blast wave evolution in
the innermost regime, $r<R_0$, (b) the intermediate phase,
corresponding to the forward and reverse shock wave evolution at
$R_0<r<R_1$, and (c) the latest phase, corresponding to blast wave
propagation at $r>R_1$.

\subsection{Emission during the early phase, $r<R_0$}

The Lorentz factor of the shocked plasma (in region $(\tilde a)$) at
$r<R_0$ is given by equation \ref{eq:Gamma_r_small}.  
The characteristic time at which radiation emitted by shocked plasma
at radius $r$ is observed by a distant observer was calculated by
\citet{W97c} for the case of an explosion into a uniform medium,
$t^{ob.} \approx r/4 \Gamma^2 c$.  
Repeating the \citet{W97c} calculation for the case of an explosion
into a density gradient $n(r) \propto r^{-2}$, we show in
\S\ref{appendix} that in this case this relation is 
slightly modified, $t^{ob.} \approx r/2 \Gamma^2 c$. 
Radiation observed at time $t^{ob.}$ is therefore emitted as the
blast wave approaches radius $r=(9E t^{ob.}/8 \pi A c)^{1/2}$. 
Radiation emitted at $r=R_0$ is observed at time
\beq
t_0^{ob.} = {8 \pi A c \over 9 E} R_0^2 \simeq 4.08 \times 10^4
\E^{-1} \ns \rs^4 {\rm \, s}, 
\label{eq:t0}
\eeq  
or about 0.47~days after the explosion.

Denoting by $\epsilon_e$ and $\epsilon_B$ the fractions of post-shock
thermal energy density, $u_{int} = 4 \Gamma^2(r) n_a(r) m_p c^2$ that
are carried by the electrons and the magnetic field respectively, the
magnetic field (in the fluid frame) is given by
\beq
\ba{lll}
B(t^{ob.}<t_0^{ob.}) & = & \left[ 8 \pi \epsilon_B \left({ 9 E \over 4 \pi}\right)
  \left( {8 \pi A c \over 9 E  t^{ob.}}\right)^{3/2} \right]^{1/2}
\nonumber \\
& = & 7.7 \times 10^{-2} \nonumber \\
& & \, \times \E^{-1/4} \eB^{1/2} {t^{ob.}_{day}}^{-3/4}
\ns^{3/4} \rs^{3/2}  {\rm \, G},
\ea 
\eeq
where $t^{ob.} = 1 {t^{ob.}_{day}}$~day.
The characteristic Lorentz factor of the shock wave accelerated
electrons is $\gamma_{char} \simeq \epsilon_e (m_p/m_e) \Gamma(r)
((p-2)/(p-1))$, where $p$ is the 
power law index of the accelerated electron energy distribution,
$n_e(\gamma) \propto \gamma^{-p}$ for $\gamma>\gamma_{char}$. The
resulting synchrotron emission peaks at 
\beq
\ba{lll}
\nu_m^{ob.}(t^{ob.}<t_0^{ob.}) & = &  5.6 \times 10^{12}(1+z)^{-1}
\nonumber \\
& & \, \times \E^{1/2} \ee^2 \eB^{1/2} {t^{ob.}_{day}}^{-3/2} \Hz, 
\ea
\eeq 
where $z$ is the redshift, characteristic values $\epsilon_e = 10^{-1}
\ee$, $\epsilon_B = 10^{-2} \eB$ are taken \citep[e.g.,][]{WG99}, and
power law index $p=2.5$ assumed.  
This frequency is below the break frequency of the spectrum, corresponding
to emission from electrons for which the synchrotron cooling time is
equal to the dynamical time, $t_{dyn} \sim r/ 8\zeta c \Gamma(r)$, 
\beq
\ba{lll}
\nu_c^{ob.}(t^{ob.}<t_0^{ob.}) & = & 7.2\times 10^{17}(1+z)^{-1}
\nonumber \\
& & \, \times \E^{1/2}
\eB^{-3/2} \zeta_0^2 {t^{ob.}_{day}}^{1/2} \ns^{-2} \rs^{-4} \Hz. 
\ea
\eeq
The synchrotron self absorption frequency is 
\beq
\ba{lll}
\nu_{\rm ssa}^{ob.}(t^{ob.}<t_0^{ob.}) & = & 7.8 \times
10^{8}(1+z)^{-1} \nonumber \\
& & \, \times 
\E^{-2/5} \ee^{-1} \eB^{1/5} \zeta_0^{-3/5}\nonumber \\
& & \, \times {t^{ob.}_{day}}^{-3/5}
\ns^{6/5} \rs^{12/5} \Hz.   
\ea
\eeq
The maximum Lorentz factor of the accelerated electrons is given by
equating the synchrotron loss time to the acceleration time, $t_{acc}
\simeq E/(cqB)$. Synchrotron emission from these electrons peaks at
\beq
\ba{lll}
\nu_{\max}^{ob.}(t^{ob.}<t_0^{ob.}) & = & 9.6\times 10^{23}(1+z)^{-1}
\nonumber \\ 
& & \, \times
\E^{1/4}{t^{ob.}_{day}}^{-1/4} \ns^{-1/4} \rs^{-1/2} \Hz.    
\ea
\eeq
Finally, the observed specific flux at $\nu_m^{ob.}$ is given by the
number of radiating electrons $N_{TOT}(r) = 4 \pi A r/m_p$ times the
observed power (per unit frequency, at $\nu_m^{ob.}$) of a single electron
$\Gamma(r)\sqrt{3}q^3 B/m_e c^2$,  
\beq
\ba{lll}
F_{\max}(t^{ob.}<t_0^{ob.}) & = & 12.7 \, \E^{1/2} \eB^{1/2} \nonumber
\\
& & \, \times
{t^{ob.}_{day}}^{-1/2} d_{L,28}^{-2} \ns \rs^2 \, {\rm mJy},    
\ea
\eeq
where $d_L = 10^{28} d_{L,28} {\rm \, cm}$ is the luminosity distance
to the GRB.

\subsection{Emission in the intermediate phase, $R_0<r<R_1$}

At $t_0^{ob.}$ the blast wave reaches $R_0$, and splits into the
relativistic forward and reverse shock waves. As we showed in
\S\ref{sec:shock_dynamics}, the Lorentz factor of the plasma is
$r$-independent in this regime. We show in appendix \ref{appendix},
that in this case the time delay suffered by photons emitted at
radius $r$ compared to photons emitted at $r=0$ is 
$\Delta t^{ob.} \approx r/ \Gamma^2(r) c$.
We thus find that this phase lasts a duration
\beq
\ba{lll}
\Delta t^{ob.} \equiv t_1^{ob.} - t_0^{ob.} & = & {R_1 - R_0 \over
  \Gamma_2^2 c} \nonumber \\
& = & 8.58 \times 10^3 \, \E^{-1} \zeta_0^{-1} \ns
\rs^4 {\rm \, s},  
\label{eq:dt}
\ea
\eeq
or about 0.10~days.

For $R_0<r<R_1$  there are 3 distinctive emitting regions: 
particles in region $(\tilde a)$ that were the only source of
emission at $r<R_0$, continue to emit; 
particles in region $(\tilde b)$ which crossed the reverse shock
and were re-heated by it; particles in region  $(\tilde c)$ that
are heated by the forward shock. All these regions are
characterized by different thermodynamic quantities, and therefore the
emission pattern in each region is unique. We thus calculate separately
emission from the different regions.

\subsubsection{Emission from particles in region $(\tilde a)$}

Region $(\tilde a)$, which was downstream of the flow past the blast
wave at $r<R_0$, becomes upstream of the flow past the reverse shock
at $r>R_0$. For $r>R_0$, the fluid can therefore only leave this
region, high energy particles are no longer injected into the region,
and no information enters it. The thermodynamic properties of
the flow (e.g., energy density, magnetic field etc.,)
``freeze out'' at their values at $r=R_0$. The values of the peak
frequency and of the spectral break frequency are therefore
equal to their values at $t^{ob.}=t_0^{ob.}$,
\beq
\ba{lll}
\nu_m^{ob.}({\tilde a}; t_0^{ob.}<t^{ob.}<t_1^{ob.}) & = & 1.8 \times
10^{13}(1+z)^{-1} \nonumber \\ 
& & \, \times \E^{2} \ee^2 \eB^{1/2} \nonumber \\ 
& & \, \times \ns^{-3/2} \rs^{-6} \Hz, 
\ea
\eeq
\beq
\ba{lll}
\nu_c^{ob.}({\tilde a}; t_0^{ob.}<t^{ob.}<t_1^{ob.}) & = &
4.9 \times10^{17}(1+z)^{-1} \nonumber \\ 
& & \, \times \eB^{-3/2} \zeta_0^2 \ns^{-3/2} \rs^{-2} \Hz. 
\ea
\eeq 
The comoving width of region $(\tilde a)$ decreases linearly from its
width at $t_0$, 
$\Delta R^{co.}_{\tilde a}(t=t_0) = c \Gamma_1 t_0^{ob.}/ 4 \zeta$ to 0
at $t_1^{ob.}$. Therefore, the optical depth decreases linearly with
time, and the synchrotron self absorption frequency is given by
\footnote{The photons have to cross several layers of plasma (regions
  $(\tilde b)$ and $(\tilde c)$) before reaching the observer. The
  observed self-absorption frequency therefore depends on the plasma
  parameters at the different regions.}  
\beq
\ba{lll}
\nu_{\rm ssa}^{ob.}({\tilde a}; t_0^{ob.}<t^{ob.}<t_1^{ob.})
& = & 1.2 \times 10^{9}\left({ t_1^{ob.} - t^{ob.} \over t_1^{ob.} -
  t_0^{ob.}} \right)^{3/5}(1+z)^{-1} \nonumber \\ 
& &  \, \times 
\E^{1/5} \ee^{-1} \eB^{1/5} \nonumber \\
& & \, \times \zeta_0^{-3/5} \ns^{3/5} \Hz.   
\ea
\eeq
Since energetic electrons are not injected into this region at
$t^{ob.} > t_0^{ob.}$, the highest energy electrons, having Lorentz
factor $\gamma_{\max}(t^{ob.}=t_0^{ob.}) = 3.2 \times 10^8 \,
\E^{-1/4} \eB^{-1/4} \rs^{3/4}$ at $t^{ob.} = t_0^{ob.}$, cool by
synchrotron radiation. Synchrotron emission from these electrons peaks
at 
\beq
\ba{lll}
\nu_{\max}^{ob.}({\tilde a}; t_0^{ob.}<t^{ob.}<t_1^{ob.}) & &
\nonumber \\
\quad =
\min\left[\nu_{\max}^{ob.}(t^{ob.}=t_0^{ob.}) \right. & , & {4.4 \times 10^{27} \over
    (t^{ob.}-t_0^{ob.})^2} (1+z)^{-1} \nonumber \\
	& & \, \times \E^{-2} \eB^{-3/2} \zeta_0^2 \nonumber \\
	& & \, \times \left.
  \ns^{1/2} \rs^{6} \Hz \right],  
\ea   
\eeq
where $\nu_{\max}^{ob.}(t^{ob.}=t_0^{ob.}) = 1.1\times 10^{24}
(1+z)^{-1}\, \E^{1/2} \ns^{-1/2} \rs^{-3/2} \Hz$ is the frequency of
the synchrotron emitted photons from the most energetic electrons at
$t^{ob.} = t_0^{ob.}$. 
At $t^{ob.}=t_1^{ob.}$,
\beq
\ba{lll}
\nu_{\max}^{ob.}({\tilde a}; t^{ob.}=t_1^{ob.}) &  =&
6.0\times 10^{19} (1+z)^{-1} \nonumber \\
& & \, \times \eB^{-3/2} \zeta_0^4  \ns^{-3/2}
\rs^{-2} \Hz.
\ea 
\eeq 
The reverse shock crosses this region at constant velocity, therefore the
number of radiating electrons in this region decreases linearly with
time. As a result, the total flux at $\nu_m^{ob.}$ decreases
linearly from its value at $t^{ob.} = t_0^{ob.}$, 
$F_{\max}({\tilde a};t^{ob.}=t_0^{ob.}) = 18.5 \, \E \eB^{1/2} d_{L,28}^{-2}
\ns^{1/2} \, {\rm mJy}$ to 0 at $t^{ob.}=t_1^{ob.}$.

\subsubsection{Emission from particles in region $(\tilde b)$}

Assuming that similar fractions $\epsilon_B = 10^{-2}$ of the post
shock thermal energy are converted to magnetic field at both the
reverse and the forward shock waves, the magnetic field produced by the
reverse shock wave 
\beq
\ba{lll}
B_{\tilde b}(prod.) & = & (8 \pi \epsilon_B u_{\tilde
  b}(t^{ob.}>t_0^{ob.}))^{1/2} 
\nonumber \\
& = & (8 \pi \epsilon_B \times 2.1
u_{\tilde a}(t^{ob.}>t_0^{ob.}))^{1/2} 
\ea
\eeq
(see eq. \ref{eq:u_tildeb}),
is smaller than the magnetic field advected with the flow from region
$(\tilde a)$,  
\beq
\ba{lll}
B_{\tilde b}(adv.) & = & {n_{\tilde b} \over n_{\tilde a}} 
B_{\tilde a}(t^{ob.}>t_0^{ob.})\nonumber \\
& = &
\sqrt{3} B_{\tilde a}(t^{ob.}>t_0^{ob.}) \nonumber \\
& = & 0.23 \, \E^{1/2} \eB^{1/2}
\rs^{-3/2} {\rm \, G}
\label{eq:B_adv}
\ea
\eeq
by a factor $\sqrt{3/2.1} \approx 1.2$.
Here, the energy density and the magnetic field in region $(\tilde a)$
assume their value at $t_0^{ob.}$ for  $t^{ob.} > t_0^{ob.}$. We
therefore assume that the magnetic field in this region is equal to
the advected magnetic field, given by equation \ref{eq:B_adv}.

The re-heating of the plasma by the reverse shock implies that the
characteristic Lorentz factor of the electrons in this region is
larger by a factor $(u_{\tilde b}/n_{\tilde b}) / (u_{\tilde
  a}/n_{\tilde a}) = 2.1/1.73 = 1.21 $ than the characteristic Lorentz
factor of electrons in region $(\tilde a)$. 
The characteristic synchrotron emission thus peaks at  
\beq
\ba{lll}
\nu_m^{ob.}({\tilde b}; t_0^{ob.}<t^{ob.}<t_1^{ob.}) & = & 3.2
\times 10^{13}(1+z)^{-1} \nonumber \\
& & \, \times \E^{2} \ee^2 \eB^{1/2} \nonumber \\
& & \, \times \ns^{-3/2} \rs^{-6} \Hz. 
\ea
\eeq

The reverse shock re-randomizes the energy, thus the spectral break
frequency is independent on the break frequency in region $(\tilde a)$,
and is given by
\beq
\ba{lll}
\nu_c^{ob.}({\tilde b}; t_0^{ob.}<t^{ob.}<t_1^{ob.}) & = & 
{8.6\times 10^{26} \over (t^{ob.}-t_0^{ob.})^2} (1+z)^{-1} \nonumber
\\ & & \, \times \E^{-2} \eB^{-3/2} \zeta_0^2 \nonumber \\ 
& & \, \times \ns^{1/2} \rs^{6} \Hz. 
\ea
\eeq 
At $t^{ob.} = t_0^{ob.}$, the comoving width of this region
is 0, hence the break frequency ${\nu_c^{ob.}}({\tilde b};
t_0^{ob.}\rightarrow t^{ob.}) \rightarrow \infty$.    
At $t^{ob.} = t_1^{ob.}$, this frequency is equal to
\beq
\ba{lll}
\nu_c^{ob.}({\tilde b}; t^{ob.}=t_1^{ob.}) & = & 
1.2\times 10^{19} (1+z)^{-1} \nonumber \\ 
& & \, \times \eB^{-3/2} \zeta_0^4 \ns^{-3/2} \rs^{-2} \Hz.
\ea
\eeq 

The self absorption frequency is
\beq
\ba{lll}
\nu_{\rm ssa}^{ob.}({\tilde b}; t_0^{ob.}<t^{ob.}<t_1^{ob.})
& = & 1.1 \times 10^{6}\left( t^{ob.} - t_0^{ob.}
\right)^{3/5}\nonumber \\
& &  \, \times 
(1+z)^{-1} \, \E^{4/5} \ee^{-1} \eB^{1/5} \nonumber \\ 
& & \, \times \zeta_0^{-3/5} \rs^{-12/5} \Hz,   
\ea
\eeq
while the highest frequency of the synchrotron emitted photons is
expected at 
\beq
\ba{lll}
\nu_{\max}^{ob.}({\tilde b}; t_0^{ob.}<t^{ob.}<t_1^{ob.}) & = &
8.5\times 10^{23} (1+z)^{-1} \nonumber \\ 
& & \, \times \E^{1/2} \ns^{-1/2} \rs^{-3/2} \Hz.      
\label{eq:e_max_b}
\ea
\eeq

The observed flux at $\nu_m^{ob.}$ emitted from this region is
expected to grow linearly from 0 at $t_0^{ob.}$ to 
\beq
\ba{lcl}
F_{\max}({\tilde b};t^{ob.}=t_1^{ob.}) & = & 
 {B_{\tilde b} \Gamma_2 \over B_{\tilde a} \Gamma_1} \times 
F_{\max}({\tilde a};t^{ob.}=t_0^{ob.}) \nonumber \\ 
& = &  \sqrt{3} \times 0.725 \times F_{\max}({\tilde
  a};t^{ob.}=t_0^{ob.}) \nonumber \\
&  = & 23.2  \, \E \eB^{1/2} d_{L,28}^{-2} \ns^{1/2} \, {\rm mJy}
\ea
\eeq
at  $t^{ob.} = t_1^{ob.}$.

\subsubsection{Emission from particles in region $(\tilde c)$}

While the energy density in region $(\tilde c)$ is equal to the energy
density in region $(\tilde b)$, there is no advected magnetic
field term. Hence, the magnetic field in this region is assumed to be
produced by the forward shock, and is equal to $B_{\tilde c} =
B_{\tilde b}(adv.)/1.2 = 1.9\times 10^{-1}  \, \E^{1/2} \eB^{1/2}
\rs^{-3/2} {\rm \, G}$ (see eq. \ref{eq:B_adv}).
Electrons are accelerated by the forward shock to characteristic
Lorentz factor $\gamma_{char}({\tilde c})
\simeq \epsilon_e (m_p/m_e) \Gamma_2 ((p-2)/(p-1))$, producing synchrotron
radiation at characteristic frequency
\beq
\ba{lll}
\nu_m^{ob.}({\tilde c}; t_0^{ob.}<t^{ob.}<t_1^{ob.}) & = & 1.0
\times 10^{13}(1+z)^{-1} \nonumber \\
& & \, \times \E^{2} \ee^2 \eB^{1/2} \nonumber \\ 
& & \, \times \ns^{-3/2} \rs^{-6} \Hz. 
\ea
\eeq 
The characteristic spectral break frequency is $\nu_c^{ob.}({\tilde
  c}) = \nu_c^{ob.}({\tilde b}) \times (B_{\tilde b}/B_{\tilde c})^3$, or
\beq
\ba{lll}
\nu_c^{ob.}({\tilde c}; t_0^{ob.}<t^{ob.}<t_1^{ob.}) & = & 
{1.5\times 10^{27} \over (t^{ob.}-t_0^{ob.})^2} (1+z)^{-1} \nonumber
\\
& & \, \times \E^{-2} \eB^{-3/2} \zeta_0^2 \ns^{1/2} \rs^{6} \Hz, 
\ea
\eeq 
which is equal to 
\beq
\ba{lll}
\nu_c^{ob.}({\tilde c}; t^{ob.}=t_1^{ob.}) & = & 
2.0\times 10^{19} (1+z)^{-1} \nonumber \\ 
& & \, \times  
\eB^{-3/2} \zeta_0^4 \ns^{-3/2} \rs^{-2} \Hz
\ea
\eeq 
at  $t^{ob.} = t_1^{ob.}$.
The synchrotron self absorption frequency is given by
\beq
\ba{lll}
\nu_{\rm ssa}^{ob.}({\tilde c}; t_0^{ob.}<t^{ob.}<t_1^{ob.})
& = &  2.5 \times 10^{6}\left( t^{ob.} - t_0^{ob.}
\right)^{3/5} \nonumber \\  
& & \, \times  (1+z)^{-1} \,
\E^{4/5} \ee^{-1} \eB^{1/5} \nonumber \\ 
& & \, \times \zeta_0^{-3/5} \rs^{-12/5} \Hz,
\ea   
\eeq
and the highest frequency of the synchrotron emitted photons is
independent of the magnetic field strength, and therefore equals to the
highest frequency of the synchrotron emitted photons in region  $(\tilde
b)$, $\nu_{\max}^{ob.}({\tilde c}; t_0^{ob.}<t^{ob.}<t_1^{ob.})
=\nu_{\max}^{ob.}({\tilde b}; t_0^{ob.}<t^{ob.}<t_1^{ob.})$ (see
eq. \ref{eq:e_max_b}). 

The number of particles swept by the forward shock from region ($b$)
into this region grows linearly with time, and is equal to 
$N_2 = 4 \pi \int_{R_0}^{R_1} r^2 n_b dr = (4/3) N_1 [(R_1/R_0)^3-1]
\approx 0.24 N_1 \zeta_0^{-1}$ at $r=R_1$ (see eq. \ref{eq:m_b}). 
Here, $N_1 = 4\pi A R_0/m_p$ is the total number of particles in
region ($a$), that were swept by the blast wave at $r=R_0$. 
The observed flux at $\nu_m^{ob.}$ emitted from this region therefore
grows linearly with time, from 0 at $t_0^{ob.}$ to 
\beq
F_{\max}({\tilde c};t^{ob.}=t_1^{ob.}) =  4.8 \, \E \eB^{1/2}
d_{L,28}^{-2} \ns^{1/2} \, {\rm mJy}, 
\eeq 
at  $t^{ob.} = t_1^{ob.}$.

\subsection{Emission in the latest phase, $r>R_1$}

As explained in \S\ref{sec:shock_dynamics}, the blast wave evolution
at $r>R_1$ is approximated by the self-similar evolution,
$\Gamma(r>R_1) = \Gamma_2 (r/R_1)^{-3/2}$. 
At $r=R_1$, the shocked shell contains matter shocked by the
reverse shock in region ($\tilde b$), as well as matter shocked by the
forward shock in region ($\tilde c$). 
While for $r>R_1$ hot matter continues to enter region ($\tilde c$)
through the forward shock wave, matter in region ($\tilde b$) cools
adiabatically. Therefore, the energy densities of the flow at the two
sides of the contact discontinuity that separates regions ($\tilde b$)
and ($\tilde c$) are different for $r>R_1$. 
For an explosion into a constant density
medium, the energy density in region ($\tilde c$) evolves as
$u_{\tilde c} \propto \Gamma^2(r) n(r) \propto r^{-3}$, while the
energy density in region ($\tilde b$) is $u_{\tilde b} \propto
r^{-4}$.
It thus follows that the contact discontinuity can not separate these
regions any more. Matter in these two regions starts to mix at
$r>R_1$, and region ($\tilde b$) eventually disappears.

We thus treat the matter in regions ($\tilde b$) and ($\tilde c$) as
having similar thermodynamic properties at $r>R_1$. This mater is
concentrated in a shell, who's observed width 
at $r=R_1$ is $\Delta R(r=R_1) = c \Delta t^{ob.} (\beta_2 - \beta_{RS}) =
0.20 \times R_1/(4 \Gamma_2^2)$. During the self similar motion that
follows, the width of the shell is $\Delta R \propto
t^{ob.}$. We therefore 
expect that the shells' width increases to its terminal value
$\Delta R(r) = r/(4 \Gamma^2(r))$ by sweeping matter into region
($\tilde c$), shortly after $t_1^{ob.}$. 
We therefore neglect emission from region ($\tilde b$) compared to
emission from region ($\tilde c$) at $r>R_1$, as well as the numerical
factor $0.20$, and calculate the emission in the latest evolutionary
phase in accordance to the self-similar solution, which is the
asymptotic solution for $r\gg R_1$.

The time delay suffered by photons emitted in this region compared to
photons emitted at $r=0$ is therefore  $\Delta t^{ob.} \approx r/4\Gamma^2(r)
c$, hence the time delay compared to photons emitted at $R_1$ is
$\Delta t^{ob.} - R_1/(4\Gamma_2^2 c)$. These photons are therefore
seen at time
\beq
t^{ob.}= t_1^{ob.} - {R_1 \over 4 \Gamma_2^2 c} + {r \over 4
  \Gamma^2(r) c}
\label{eq:t_hat}
\eeq 
from the explosion. This equation is written in the form
\beq
\tilde{t^{ob.}} = {r \over 4 \Gamma^2 c}, 
\eeq
where $\tilde{t^{ob.}} \equiv t^{ob.} -t_1^{ob.} + R_1/(4 \Gamma_2^2
c)$.

Using this scaling, the magnetic field is given by
\beq
\ba{lll}
B(t^{ob.}>t_1^{ob.}) & = & \left(8 \pi \epsilon_B \times 4 \Gamma^2 n_b
m_p c^2 \right)^{1/2} \nonumber \\ 
& = & 0.15 \, \E^{1/8} \eB^{1/2}
{\tilde{t^{ob.}_{day}}}^{-3/8} \ns^{3/8} {\rm \, G},  
\ea
\eeq
where $\tilde{t^{ob.}} = 1 \tilde{t^{ob.}_{day}}$~day. 
Synchrotron radiation therefore peaks at 
\beq
\ba{lll}
\nu_m^{ob.}(t^{ob.}>t_1^{ob.}) & = & 3.5 \times 10^{12}(1+z)^{-1}
\nonumber \\ 
& & \, \times \E^{1/2} \ee^2 \eB^{1/2} {\tilde{t^{ob.}_{day}}}^{-3/2} \Hz. 
\ea
\eeq
The break frequency in the spectrum is at  
\beq
\ba{lll}
\nu_c^{ob.}(t^{ob.}>t_1^{ob.}) & = &  3.2\times 10^{16}(1+z)^{-1} \nonumber
\\ 
& & \, \times \E^{-1/2}
\eB^{-3/2} \zeta_0^2 {\tilde{t^{ob.}_{day}}}^{-1/2} \ns^{-1} \Hz, 
\ea
\eeq
and the self absorption frequency is 
\beq
\ba{lll}
\nu_{\rm ssa}^{ob.}(t^{ob.}>t_1^{ob.}) & = & 3.5 \times
10^{9}(1+z)^{-1} \nonumber \\ 
& & \, \times 
\E^{1/5} \ee^{-1} \eB^{1/5} \zeta_0^{-3/5} \ns^{3/5} \Hz.   
\ea
\eeq
The highest frequency of the synchrotron emitted photons is expected at 
\beq
\ba{lll}
\nu_{\max}^{ob.}(t^{ob.}>t_1^{ob.}) & = &  6.5\times 10^{23}(1+z)^{-1}
\nonumber \\ 
& & \, \times 
\E^{1/8} {\tilde{t^{ob.}_{day}}}^{-3/8} \ns^{-1/8} \Hz.   
\ea
\eeq
The total number of radiating electrons at radius $r \gg R_1$ is approximated
by $N(r) = (4/3) N_1 [(r/R_0)^3-1] \approx (4/3) N_1 (r/R_0)^3$, thus
the asymptotic value of the flux emitted at $\nu_m^{ob.}$ is given by
\beq
F_{\max}(t^{ob.}>t_1^{ob.}) = 38.7 \, \E \eB^{1/2}
d_{L,28}^{-2} \ns^{1/2} \, {\rm mJy}    
\eeq 
(a somewhat lower value is expected at $r \gtrsim R_1$, with no simple
dependence on the values of the unknown parameters).

\subsection{Predicted afterglow light curves}

Light curves are calculated using standard formulae for the flux at
 a given frequency \citep[e.g.,][]{SPN98}, 
\beq
F_\nu = \left\{ \begin{array}{ll}
F_{\max} (\nu_{ssa}/\nu_m)^{1/3}(\nu/\nu_{ssa})^2 & \nu <
\nu_{ssa}, \\
F_{\max} (\nu/\nu_m)^{1/3} & \nu_{ssa} < \nu < \nu_m, \\
F_{\max} (\nu/\nu_m)^{-(p-1)/2} & \nu_m < \nu < \nu_c, \\
F_{\max} (\nu_c/\nu_m)^{-(p-1)/2} (\nu/\nu_c)^{-p/2} & \nu_c < \nu < \nu_{\max}.
\end{array}\right.
\eeq
The calculation is done separately for each of the 3 different
phases. In the intermediate phase $t_0^{ob.} < t^{ob.} < t_1^{ob.}$,
emission from the 3 different emitting regions is calculated
separately and summed up. 

\begin{figure}          
\plotone{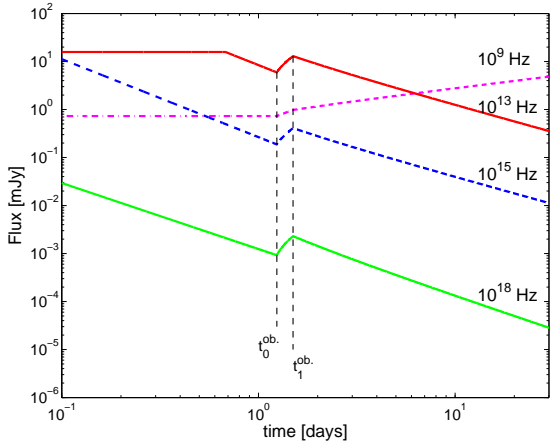}          
\caption{Schematic description of the light curves at different
  frequencies, Radio ($10^9$~Hz), mid-IR ($10^{13}$~Hz), Optic
  ($10^{15}$~Hz) and X-ray ($10^{18}$~Hz). Power law index $p=2.5$ of
  the accelerated electrons assumed, and luminosity distance $d_L =
  10^{28}{\rm \, cm}$ assumed in producing this plot. 
}          
\label{fig:Lightcurve}          
\end{figure}

The resulting light curves for radio ($10^9$~Hz), mid-IR
($10^{13}$~Hz), optical ($10^{15}$~Hz) and X-ray ($10^{18}$~Hz)
frequencies are presented in Figure~\ref{fig:Lightcurve}.
In producing this figure, characteristic values of $\rs = 1.6$,
$\ns=0.4$ were assumed, which result in $t_0^{ob.} \approx 0.6$~day
and $\Delta t^{ob.} = 0.13$~day (see eqs. \ref{eq:R_0},\ref{eq:n_a},
\ref{eq:t0}, \ref{eq:dt}).  
The radio band is at the characteristic frequency of the synchrotron self
absorption frequency at all times, the mid-IR frequency is close to the
peak emission, the optical band is typically higher than $\nu_m^{ob.}$ and
lower than $\nu_c^{ob.}$, while the X-ray frequency is above $\nu_c^{ob.}$ for
the characteristic parameters chosen.

Two important features of this scenario are seen in this figure. 
The first, which is common to all frequencies, is the increase of the
flux by a factor $\gtrsim 2$ from $t_0^{ob.}$ to $t_1^{ob.}$ caused by
the simultaneous radiation from the 3 different regions during this period.  
The second feature is the flattening of the power law index $\alpha$ of the
flux time dependence ($F_\nu \propto t^{-\alpha}$, where $\alpha =
\alpha(\nu)$ is frequency dependent) from its value at
$t^{ob.}<t_0^{ob.}$ to a  value 
smaller by $1/2$ at late times, $t^{ob.}>t_1^{ob.}$. 
This flattening is caused by the change of the ambient density
profile, from $n(r) \propto r^{-2}$ at $t^{ob.}<t_0^{ob.}$ to constant
density profile, $n(r) \propto r^0$ at late times of the blast wave
evolution. 
The power low index therefore changes accordingly:
for $\nu<\nu_{ssa}$, $F_\nu\propto t^1$ if $n(r) \propto
r^{-2}$, and $F_\nu\propto t^{1/2}$ if $n(r) \propto r^0$;
 for $\nu_{ssa}<\nu<\nu_m$, $F_\nu\propto t^0$ if $n(r) \propto
r^{-2}$, and $F_\nu\propto t^{-1/2}$ if $n(r) \propto r^0$; 
and for $\nu_m<\nu<\nu_c$, $F_\nu\propto t^{-(3p-1)/4}$ if $n(r) \propto
r^{-2}$, and $F_\nu\propto t^{-(3p-3)/4}$ if $n(r) \propto r^0$.
At high frequencies, the light curve time dependence is not changed by
the change of the density profile, $F_\nu\propto t^{-(3p-2)/4}$ for
$\nu>\nu_c$.

\section{Summary and discussion}
\label{sec:summary}

In this work, we considered the scenario of a massive star as a GRB
progenitor. This scenario results in a complex structure of the
circumburst medium, which is composed of 4 different regions:
unshocked wind, shocked wind, shocked ISM and unshocked ISM.
We showed that the main effect takes place when the blast wave reaches
the wind-reverse shock which separates the unshocked wind and the
shocked wind regions. The blast wave then splits, and a blast wave
reverse shock is formed. We showed in \S\ref{sec:shock_collision} that
the blast wave reverse shock, which separates two hot regions is not
strong. 
By solving the equations describing the shock jump conditions, we
showed that as long as the reverse shock exists, the shocked plasma
moves at constant velocity, and simple analytic relation between the
Lorentz factors of the flow at the different regimes exists
(eq. \ref{eq:Gamma2}). 
We then calculated the blast wave evolution, and the resulting light
curves at different frequencies 
(\S\ref{sec:lightcurve}, fig. \ref{fig:Lightcurve}). 
   
The resulting light curves are significantly different than
``standard'' afterglow light curves calculations, which assume
explosion into uniform medium, or into a density gradient. 
The resulting light curve in the scenario presented here
is similar to the light curve of an explosion into a density gradient at
early times before the flux rise, and to light curves obtained for an
explosion into a uniform medium at late times. It has an
important additional feature: during a short transition, lasting
$\Delta t^{ob.} / t_0^{ob.} \simeq 1/5$ of the transition time, which 
is expected to take place at $\simeq 1$~day after the GRB explosion, the
flux rises by a factor of $\gtrsim 2$ at all frequencies.  
Such a rise in the flux is a prediction of this model. This model may
therefore provide a natural explanation to the rise in the various
optical bands fluxes of GRB030329 by a factor of $4$ observed after $\sim
1.4$ days \citep{L04}.  

The occurrence time of the transition $t_0^{ob.}$,
is the time at which radiation that was emitted as the blast wave
reached the radius of the wind reverse shock $R_0$ is observed.
This time is weakly constrained, being strongly dependent on the
unknown values of $R_0$ and of the wind density at this radius,
$n_{R_0}$. Therefore characteristic transition occurrence times lasting
between few hours to several days may be expected. 
The transition time $\Delta t^{ob.}$ 
on the other hand, have the same parametric dependence as $t_0^{ob.}$.
We therefore expect longer rise of the flux in bursts for which the
transition occurs at late times, compared to bursts for which the
transition occurs at earlier times.  

The purely constrained value of the transition time $t_0^{ob.}$
implies that observed afterglow light curves from different bursts 
may be interpreted as resulting from explosion into uniform medium, or
into density gradient, if only part of the data is available. Thus, we
find that both interpretations \citep[e.g.,][]{CL00} may be consistent
with the data, with preferably parameters that fit explosion into
uniform medium at very late times.

We used here a highly simplified model to describe the density profile
of the ambient matter. We did not consider radiative cooling, which is
highly uncertain and, if exists, lowers the temperature of the shocked
wind gas in region $(b)$, hence rises its density. We also did not
consider inhomogeneities inside the different regions, which may also
have observational consequences.  
The calculation
presented in \S\ref{sec:shock_collision} for the interaction of a
relativistic blast wave with a density discontinuity caused by a
strong shock is, however, general, and can be used in the context of
supernovae calculations as well. We thus conclude that the main
findings of this work - the afterglow re-brightening by a factor of
$\gtrsim 2$, and the different light curves slopes at early and late
times, remain valid also in a more complex and realistic ambient medium
profile.

\acknowledgments
AP wishes to thank James Miller-Jones and Peter M\'esz\'aros for
useful discussions. 
This research was supported by NWO grant 639.043.302 to RW and by the
EU under RTN grant HPRN-CT-2002-00294.

\appendix
\section{Relation between the observed time and emission radius in
  relativistic fireball}
\label{appendix}

\citet{W97c} calculated the relation between the observed time delay
$\Delta t^{ob.}$ of photons emitted by synchrotron
radiation from a shell expanding relativistically and self-similarly
in a uniform medium as the shell's front reaches radius $r$, with
respect to photons emitted from the center of the explosion at $r=0$,
and found the relation $\Delta t^{ob.} \simeq r/(4\Gamma^2(r) c)$.  
Here, we repeat the calculation for the case of a relativistic
expansion in a density gradient, $n(r)\propto r^{-p}$. 
In this case, assuming adiabatic expansion, the Lorentz factor of the
flow scales with the radius as $\Gamma(r)\propto r^{-(3-p)/2}$. The
magnetic field scales as $B\propto \Gamma(r) n(r)^{1/2}\propto
r^{-3/2}$, and the characteristic Lorentz factor of the electrons,
$\gamma_{char} \propto \Gamma(r)$. 
The peak frequency of the synchrotron emitted photons (in the comoving
frame) therefore scales as $\nu_m \propto \gamma_{char}^2 B \propto
r^{p-9/2}$. 

The number of photons emitted by a single electron at a
unit comoving time $\dot N_\gamma$ is proportional to the magnetic field $B$. 
The total number of electrons swept by the shell (and,
presumably, emit) at radius $r$ is $N_e(r) \propto r^{3-p}$.   
Since the comoving time during which the plasma propagates a distance
$dr$ is $dt^{co.} \simeq dr/\Gamma(r)c$, the number of photons emitted
as the plasma expands a distance $dr$ scales as  
$dN_\gamma/dr \propto N_e {\dot N_\gamma} (dt^{co.}/dr) \propto
r^{3-3p/2}$. 
The fraction of photons emitted at frequency $\nu_m$ that are observed
at frequency $\nu_m^{ob.}$ is given by eq. 5 of \citet{W97c},
$df/d\nu_m^{ob.} \simeq (2 \Gamma \nu_m)^{-1}$, where $\Gamma\gg 1$
assumed. Thus, the number of photons produced by the fireball shell at
radius $r$ with frequency in the range $\nu_m^{ob.} .. \nu_m^{ob.}
+d\nu_m^{ob.}$ is 
\beq
{d^2N_\gamma \over d\nu_m^{ob.} dr} \propto {dN_\gamma\over dr}
\Gamma(r)^{-1} \nu_m^{-1} \propto r^{3(3-p)}.
\eeq

The delay in observed time of a photon emitted from the edge of the
shell at frequency $\nu_m$ and observed at frequency $\nu_m^{ob.}$,
with respect to photons emitted on the line of sight is given by (for
$\Gamma \gg 1$) 
\beq
\Delta t_\theta(\nu_m^{ob.},r) = {r \over \Gamma^2 c}\left( \Gamma {\nu_m \over
  \nu_m^{ob.}} - {1 \over 2} \right) \propto r^{4-p}
  \left(r^{-6+3p/2} -1/2 \right).
\eeq 
An additional delay exists between photons emitted at radius $r$ on
the line of sight, compared to photons emitted from the center of the
explosion at $r=0$,
\beq
\Delta t_r(r) = {r \over c} - \int_0^r {dr \over \left(1-{1
    \over2\Gamma^2(r)}\right)^{1/2} c} = {r \over 4(4-p) \Gamma^2(r) c}.
\eeq

Assuming that photons are emitted uniformly from a shell of finite
thickness $\xi r/\Gamma^2 c$, the arrival time of photons emitted at
radius $r$ and observed with frequency $\nu_m^{ob.}$ are uniformly
distributed over the range $t(\nu_m^{ob.},r) = \Delta
t_\theta(\nu_m^{ob.},r) + \Delta t_r(r)$, and  $t(\nu_m^{ob.},r) +
\tau_\xi$, where $\tau_\xi = \xi r/\Gamma^2 c$. 

\begin{figure}          
\plotone{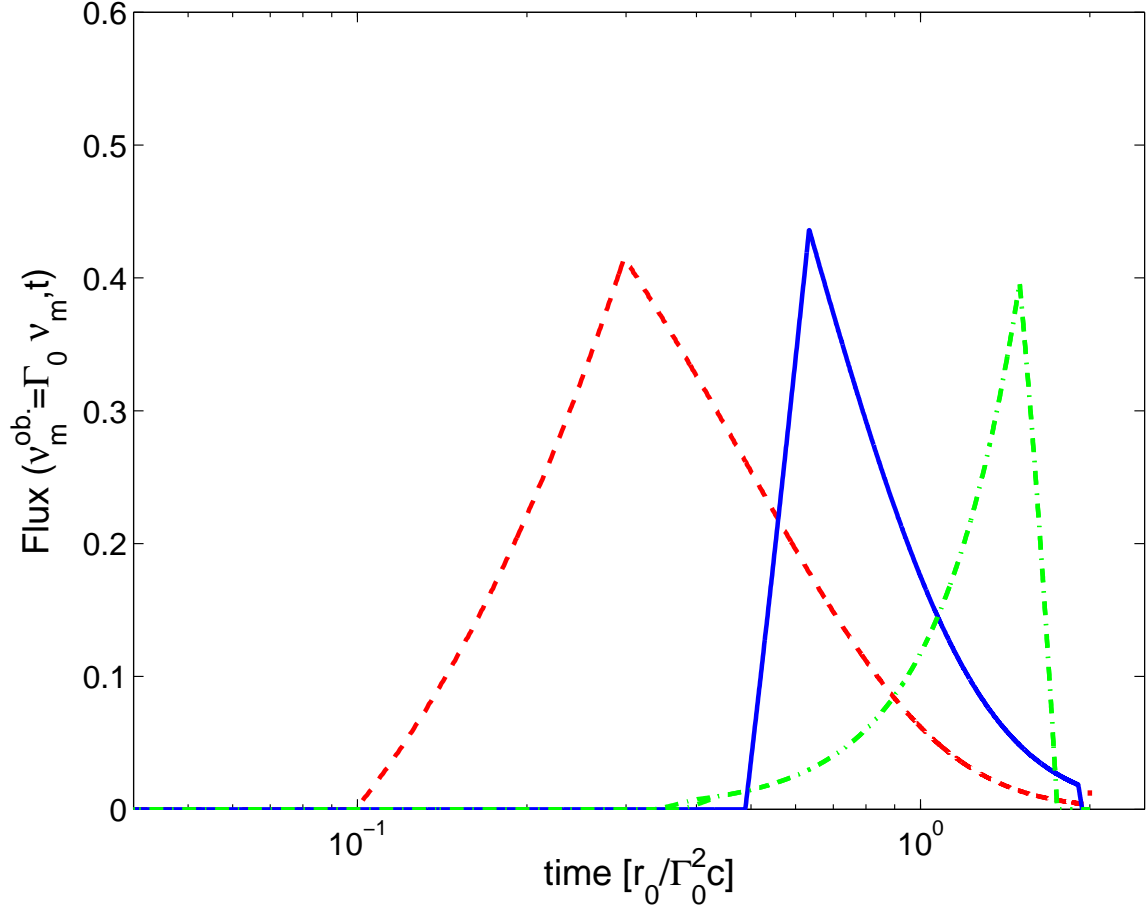}          
\caption{
Normalized observed flux at observed frequency $\nu_m$ as a function
of time, for various density profile considered. The solid line shows
the result for a density profile $n(r) \propto r^{-2}$ as expected in
the early stages of the blast wave evolution, at $r<R_0$. The dash
line presents the results for an expansion in homogeneous medium, and
is similar to the results of \citep{W97c}. The dash-dotted line
represents a scenario of $r$-independent Lorentz factor and magnetic
field, as expected for $R_0<r<R_1$. 
}          
\label{fig:Appendix}          
\end{figure}

Figure \ref{fig:Appendix} shows the results of a numerical calculation
of the flux as a function of time \citep[compare with fig. 1 of
][]{W97c}. The width of the shell considered is $\xi=1/8$. The three
plots represent the three cases considered in this work. Explosion
into density gradient, $n(r)\propto r^{-2}$ is presented by the solid
curve, and explosion into uniform medium is presented by the dashed
curve. The Dash-dotted curve represents the special scenario that
exists in the intermediate phase of the blast wave evolution at
$R_0<r<R_1$, during which the Lorentz factor and the 
magnetic field are kept constant as long as the reverse shock exists. 

We thus conclude that while for an expansion into uniform medium,
\citet{W97c} relation, $t^{ob.} \simeq r/4 \Gamma^2 c$ holds, for an
explosion into density gradient,  $n(r)\propto r^{-2}$ this relation
is  modified, and is $t^{ob.} \simeq r/2 \Gamma^2 c$, while in the
intermediate phase of the blast wave evolution the correct relation is
$t^{ob.} \simeq r/ \Gamma^2 c$.

\end{document}